\documentclass[aps,twocolumn,showpacs,superscriptaddress,groupedaddress,nofootinbib]{revtex4} 
\usepackage{graphicx}
\usepackage[sort&compress]{natbib}
\usepackage{subfigure}
\usepackage{amsmath}
\usepackage{amssymb}
\usepackage{amsfonts}
\usepackage{rotating}
\usepackage{cancel}
\usepackage{lipsum}
\usepackage{mathtools}
\usepackage{color}
\usepackage{bbm}
\usepackage{dsfont}
\usepackage{bbold}
\usepackage{multirow}
\usepackage{ulem}
\newcommand{\eps}{\epsilon}
\begin{document}

\title{Stripping triangle loops: Discussion of $D_s^+\to \rho^+\eta\to \pi^+\pi^0\eta$ in $a_0(980)$ production}
\author{M.~Bayar}
\email{melahat.bayar@kocaeli.edu}
\affiliation{Department of Physics, Kocaeli Univeristy, 41380, Izmit, Turkey}  
\author{R.~Molina}
\email{Raquel.Molina@ific.uv.es}
\affiliation{Departamento de F\'{\i}sica Te\'orica and IFIC,
Centro Mixto Universidad de Valencia-CSIC, Parc Científic UV, C/ Catedrático José Beltrán, 2, 46980 Paterna, Spain} 
\author{E.~Oset}
\email{oset@ific.uv.es}
\affiliation{Departamento de F\'{\i}sica Te\'orica and IFIC,
Centro Mixto Universidad de Valencia-CSIC, Parc Científic UV, C/ Catedrático José Beltrán, 2, 46980 Paterna, Spain} 
\author{Ming-Zhu Liu}
\affiliation{ School of Nuclear Science and Technology, Lanzhou University, Lanzhou 730000, China}

\author{Li-Sheng Geng}\email{ lisheng.geng@buaa.edu.cn}
\affiliation{School of Physics, Beihang University, Beijing 102206, China}
\affiliation{Beijing Key Laboratory of Advanced Nuclear Materials and Physics, Beihang University, Beijing 102206, China}
\affiliation{Peng Huanwu Collaborative Center for Research and Education, Beihang University, Beijing 100191, China}
\affiliation{Southern Center for Nuclear-Science Theory (SCNT), Institute of Modern Physics, Chinese Academy of Sciences, Huizhou 516000, China}

\begin{abstract}  
We address a general problem in the evaluation of triangle loops stemming from the consideration of the range of the interaction involved in some of the vertices, as well as the energy dependence of the width of some unstable particles in the loop. We find sizeable corrections from both effects. We apply that to a loop relevant to the $D_s^+ \to \pi^+ \pi^0 \eta $ decay, and find reductions of about a factor of $4$ in the mass distribution of invariant mass of the $\pi \eta$ in the region of the $a_0(980)$. The method used is based on the explicit analytical evaluation of the $q^0$ integration in the $d^4q$ loop integration, using Cauchy's residues method, which at the same time offers an insight on the convergence of the integrals and the effect of form factors and cutoffs. 

\end{abstract}
\maketitle
\section{Introduction}
Triangle loops, with three intermediate propagators, are often encountered in the study of hadronic processes. A peculiar type of triangle loops are those that develop a triangle singularity \cite{karplus,landau} in which the three intermediate particles can be placed simultaneously on shell, are collinear and the process can occur at the classical level (Coleman Norton Theorem \cite{coleman}). A recent review of this issue is given in \cite{guosakai} and a practical method to see when a diagram develops a triangle singularity is given in \cite{trianus}. 

An interesting case of triangle mechanism was discussed in \cite{hsiao} and \cite{lingeng} for the interpretation of the $D_s^+\to \pi^+\pi^0\eta $ reaction measured by the BESIII collaboration \cite{besexpe}. In the experiment of \cite{besexpe} it was claimed that the reaction proceeds via quark annhilation with an abnormally large rate for that decay mode. However, both in \cite{hsiao} and \cite{lingeng} it was shown that the process could be obtained via the triangle mechanism in which the $D_s^+$ decays to $\rho^+\eta$,  proceeds via external emission and is Cabibbo favored. The $\rho^+$ decays further into $\pi^+\pi^0$ and $\pi^+\eta$ or $\pi^0\eta$ produces the $a_0(980)$ through rescattering of these mesons, which is the origin of the $a_0(980)$ in the chiral unitary approach \cite{ollernpa,norbert,loher,juanito}. Prior to these theoretical works, the mechanism of internal emission was shown to give a perfect reproduction of the experimental data, although the absolute branching ratio could not be evaluated there. 

One difference between the work of \cite{hsiao} and \cite{lingeng} is the use of the $\pi\eta\to\pi\eta$ amplitude from the chiral unitary approach in \cite{lingeng}, which gives rise to the $a_0$ effective propagator, while an empirical $a_0$ propagator is used in \cite{hsiao}. In addition, it was also shown in \cite{lingeng} that the triangle mechanism $D_s^+\to K^*\bar{K}\to K\pi\bar{K}$, followed by rescattering of $K\bar{K}$ to produce the $a_0(980)$, $K\bar{K}$ being one of the coupled channels considered in the chiral unitary approach to produce the $a_0(980)$, is also important in the process and even has larger strength than the triangle loop through $\rho^+$ decay.

In both works of \cite{hsiao} and \cite{lingeng}, fourfold $\int d^4 q$ loop integrations were carried out using tools of Feynmann integrals. Our purpose here is to show that when dealing with triangle loops that involve some strong interaction, like $\pi\eta\to \pi\eta$ or $K\bar{K}\to \pi\eta$, the range of the interaction introduces corrections in the loops resulting in sizeable reductions of the loop strength. The implementation of these corrections is easily done if one performs the $q^0$ integration analytically using Cauchy integration. The procedure also allows to take into account corrections stemming from the energy dependence of the $\rho$ width in the loop (a constant $\Gamma_\rho$ is used in \cite{hsiao} and \cite{lingeng}). The purpose of the work is not to get a new picture for the $D_s^+\to \pi^+\pi^0\eta$ decay, but to show a technical method to evaluate accurately triangle loops involving a strong hadron-hadron interaction in some vertices. Concerning the $D_s^+\to \pi^+\pi^0\eta$ reaction, the reduction that we find in the triangle mechanisms, indicates the relevance of the findings of \cite{raquelgeng} with the internal emission mechanism, stressing the conclusion of \cite{lingeng}: The triangle loop mechanism from external emission, and the mechanism of \cite{raquelgeng} with internal emission are both at work in the $D_s^+\to \eta\pi^+\pi^0$ and one does not need the quark annihilation mechanism. 
\section{Formalism}
\begin{figure}
 \centering
 \includegraphics[scale=0.72]{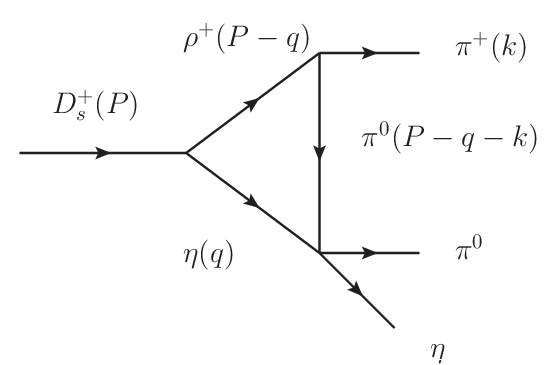}
 \caption{Triangle loop diagram for the $D_s^+$ decay to $\pi^+\pi^0\eta$. The momenta of the particles are shown in parenthesis.}
 \label{fig:decay}
\end{figure}

The loop that we are considering is given in Fig.~\ref{fig:decay}. 

In \cite{lingeng}, the detailed structure of the weak $D_s^+$ decay to $\rho^+\eta$ is taken into account including form factors for the different terms. Since our purpose is to show the effect of the range of the strong interaction in the loop, we find sufficient to take a $p$-wave vertex for the $D_s^+\to\rho^+\eta$ decay, and a form factor associated to the momentum $q$ in Fig.~\ref{fig:decay}. One could take the form factor associated to the $\rho$ momentum with the same conclusions. We take a $D_s^+\to \rho^+\eta$ vertex of the type,
\begin{equation}
 t_{D_s,\rho\eta}\equiv C\eps_\mu(\rho)(P+q)^\mu\ ,
 \label{eq:ver}
\end{equation}
and a form factor from \cite{lingeng}
\begin{equation}
 F(q)=\frac{1}{1-a(q^2/m_{D_s}^2)+b(q^4/m_{D_s}^4)}\ ,
 \label{eq:ff}
\end{equation}
with $a=0.69$ and $b=0.002$. The $\rho^+$ decay to $\pi^+\pi^0$ can be obtained from the standard lagrangian,
\begin{eqnarray}
 {\cal L}=-ig\langle{\left[P,\partial_\mu P\right]V^\mu\rangle}\ ,
\end{eqnarray}
with $g=\frac{M_V}{2f}$, $M=800$~MeV, $f=93$~MeV, where $P,V$, are the pseudoscalar and vector SU(3) matrices, respectively, representing the $q\bar{q}$ written in terms of mesons~\cite{vini}. The $\rho^+\to \pi^+\pi^0$ decay vertex is given by,
\begin{eqnarray}
 -it&=&-ig\sqrt{2}\eps^\mu(\rho)(p_{\pi^+}-p_{\pi^0})_\mu\nonumber\\&=&-ig\sqrt{2}\eps^\mu(\rho)(2k+q-P)_\mu\ .
\end{eqnarray}
The loop function corresponding to Fig.~\ref{fig:decay} is readily written as,
\begin{widetext}
\begin{eqnarray}
 -it=Cg\sqrt{2}t_{\pi^0\eta,\pi^0\eta}(M_{\mathrm{inv}})\int\frac{d^4q}{(2\pi)^4}   \frac{\eps_\mu(\rho)(P+q)^\mu \eps_\nu(\rho)(2k+q-P)^\nu }{(q^2-m^2_\eta+i\eps)((P-q)^2-m^2_\rho+i\eps)((P-q-k)^2-m_{\pi^0}^2+i\eps)}\ ,
 \label{eq:tmat}
\end{eqnarray}
\end{widetext}
where, since $M_{\mathrm{inv}}$ is the external $\pi^0\eta$ invariant mass, $t_{\pi^0\eta,\pi^0\eta}$, can be taken out of the integral. Then, summing over the $\rho$ polarizations,
\begin{eqnarray}
 \sum_{\mathrm{pol}}\eps_\mu(\rho)\eps_\nu(\rho)=-g_{\mu\nu}+\frac{(P-q)_\mu(P-q)_\nu}{M^2_\rho}\ ,
\end{eqnarray}
we obtain,
\begin{widetext}
\begin{eqnarray}
 t=&&iCg\sqrt{2}\,t_{\pi^0\eta,\pi^0\eta}(M_{\mathrm{inv}})\int\frac{d^4q}{(2\pi)^4}\frac{1}{(q^2-m^2_\eta+i\eps)((P-q)^2-m_\rho^2+i\eps)((P-q-k)^2-m_{\pi^0}^2+i\eps)}\nonumber\\&\times &
 \left\{-(P^0+q^0)(2k^0+q^0-P^0)+\vec{q}\cdot(2\vec{k}+\vec{q}\,)+\frac{1}{M^2_\rho}(P^2-q^2)[(P^0-q^0)(2k^0+q^0-P^0)+\vec{q}\cdot(2\vec{k}+\vec{q}\,)]\right\}\ .\label{eq:loop}
\end{eqnarray}
\end{widetext}
We should implement the form factor of Eq.~(\ref{eq:ff}) there. These form factors in weak interactions are meant to introduce some $q$ dependence when the $\rho^+\eta$ particles are on shell. The eventual poles generated by them are not physical. With this view, we shall consider the $F(q)$ factor for the values of $q^2$ obtained with the Cauchy integration with the structure of the three propagators in Eq.~(\ref{eq:loop}). 

There is another factor to take into consideration. In Eq.~(\ref{eq:loop}), we have factorized the $t_{\pi^0\eta\to\pi^0\eta}$ amplitude on-shell outside of the integrand, when we should have used the half off shell amplitude. It is shown in \cite{danijuan} that the unitary approach with the $G$ function regularized with a sharp cutoff in the three momentum is equivalent to the use of a potential,
\begin{eqnarray}
 V(\vec{q},\vec{q}\,')=V\theta(q_\mathrm{max}-|\vec{q}\,|)\theta(q_\mathrm{max}-|\vec{q}\,'|)\ ,\label{eq:co}
\end{eqnarray}
which automatically reverts into,
\begin{eqnarray}
 T(q,q')=T\theta(q_\mathrm{max}-|\vec{q}\,|))\theta(q_\mathrm{max}-|\vec{q}\,'|)\ ,\label{eq:tco}
\end{eqnarray}
where $q,q'$ are the initial or final momenta in the rest frame of $\pi\eta$ in the present case. In this way, the $t_{\pi^0\eta,\pi^0\eta}$ matrix in Eq.~(\ref{eq:loop}) contains a factor inside of the loop, $\theta(q_\mathrm{max}-|\vec{q}\,^\star|)$, with $\vec{q}\,^\star$ the momentum $\vec{q}$ in the rest frame of $\pi^0\eta$. This is given by
\begin{eqnarray}
 \vec{q}\,^\star=\left[\left(\frac{E_R}{M_\mathrm{inv}}-1\right)\frac{\vec{q}\cdot\vec{k}}{\vec{k}\,^2}+\frac{q^0}{M_\mathrm{inv}}\right]\vec{k}+\vec{q}\ ,\label{eq:boost}
\end{eqnarray}
with $M_R=\sqrt{M_\mathrm{inv}^2+\vec{k}\,^2}$, and $q^0$ given by the value of this variable in the Cauchy $q^0$ integration by residues. The same can be said about the value of $q^2=q^{0\,2}-\vec{q}\,^2$ in the form factor of Eq.~(\ref{eq:ff}). 

Let us then perform the Cauchy $q^0$ integration analytically. For this, we need to know the pole structure of the three propagators. At this point, let us write any meson propagator as,
\begin{eqnarray}
 \frac{1}{q^2-m^2}=\frac{1}{2\omega(q)}\left(\frac{1}{q^0-\omega(q)+i\eps}-\frac{1}{q^0+\omega(q)-i\eps}\right)\ .\nonumber\\\label{eq:prop}
\end{eqnarray}
The first term in Eq.~(\ref{eq:prop}) is the one of positive energy, and the second corresponds to the negative energy part.

In the process of Fig.~\ref{fig:decay}, the $D_s^+$ can decay physically to $\rho^+\eta$. With the direction of the arrows in the figure, the $\eta$ propagator is largely dominated by the positive energy part of the propagator which develops a pole at $q^0=\omega(q)$, the on-shell $\eta$ energy in the physical $D^+_s$ decay to $\rho^+\eta$. The same can be said about the $\rho$ propagator. However, we cannot say that about the internal $\pi^0$ propagator in the loop. In the $a_0(980)$ region for $\pi^0\eta$, the $\pi^+$ carries $771$~MeV of energy, which leaves little energy for the $\pi^0$, which is largely off-shell. Then, for the $\pi^0$ in the loop, we keep the two terms of the propagator of Eq.~(\ref{eq:prop}). Performing the Cauchy integration on the semicircle above or below the real axis, we  find,
\begin{equation}
 t=t_1+t_2\ ,\label{eq:t}
\end{equation}
where $t_1$ corresponds to the $[2\omega(q^0-\omega+i\eps)]^-1$ part of the $\pi^0$ propagator, and $t_2$ to the $-[2\omega(q^0+\omega-i\eps)]^-1$ part. We then find the residues at $q_0=\omega_\eta(q)$, for $t_1$, and at $q^0=P^0-\omega_\rho(q)$ for $t_2$, and,
\begin{widetext}
\begin{eqnarray}
 &&t_1=C\sqrt{2}g t_{\pi^0\eta,\pi^0\eta}(M_\mathrm{inv})\int \frac{d^3q}{(2\pi)^3}\frac{\theta(q_\mathrm{max}-|\vec{q}\,^\star|)\,{\mathcal P}_1\,F(q^2)}{8\omega_\eta(q)\omega_\rho(q)\omega_\pi(\vec{k}+\vec{q}\,)(P^0-\omega_\eta(q)-\omega_\rho(q)+i\eps)(P^0-\omega_\eta(q)-k^0-\omega_\pi(\vec{k}+\vec{q}\,)+i\eps)}\nonumber\\\label{eq:t1}
 &&t_2=C\sqrt{2}g t_{\pi^0\eta,\pi^0\eta}(M_\mathrm{inv})\int \frac{d^3q}{(2\pi)^3}\frac{\theta(q_\mathrm{max}-|\vec{q}\,^\star|)\,{\mathcal P}_2\,F(q^2)}{8\omega_\eta(q)\omega_\rho(q)\omega_\pi(\vec{k}+\vec{q}\,)(P^0-\omega_\eta(q)-\omega_\rho(q)+i\eps)(k^0-\omega_\rho(q)-\omega_\pi(\vec{k}+\vec{q}\,)+i\eps)}\ ,\nonumber\\\label{eq:t2}
\end{eqnarray}
\end{widetext}
where ${\mathcal P}_1$ and ${\mathcal P}_2$, are given by,
\begin{widetext}
\begin{eqnarray}
 &&{\mathcal P}_1=M_{D_s}^2-m_\eta^2+2\vec{k}\cdot\vec{q}-2k^0(M_{D_s}+\omega_\eta(q))+\frac{1}{M^2_\rho}(M^2_{D_s}-m_\eta^2)[2k^0(P^0-\omega_\eta(q))-M^2_{D_s}-m^2_\eta+2P^0\omega_\eta(q)+2\vec{k}\cdot\vec{q}\,]\nonumber\\
 &&{\mathcal P}_2=2M_{D_s}\omega_\rho(q)-m_\rho^2-2k^0(2M_{D_s}-\omega_\rho(q))+2\vec{k}\cdot\vec{q}+\frac{1}{M^2_\rho}(2 M_{D_s}\omega_\rho(q)-m_\rho^2)(2k^0\omega_\rho(q)-m_\rho^2+2\vec{k}\cdot\vec{q}\,)\ .\label{eq:ps}
\end{eqnarray}
\end{widetext}
Note that $F(q^2)$ in $t_1$ is $F(m^2_\eta)$, but $F(q^2)$ in $t_2$ is $F(P^0-\omega_\rho(q))^2-\vec{q}\,^2)=F(M^2_{D_s}+m_\rho^2-2M_{D_S}\omega_\rho(q))$. Hence, the form factor does not reduce the degree of divergence of the loop of $t_1$ but it does in the loop of $t_2$. Interestingly, we see that, omitting the $\theta(\cdot)$ factor, $t_1$ is convergent, but $t_2$ is logarithmically divergent. The form factor $F(q^2)$ in this case renders the terms convergent, as a normal counting in the $\int d^4 q$ integration would have given us. Yet, the factor $\theta(q_\mathrm{max}-|\vec{q}\,^\star|)$ will reduce both loop functions and we shall see how this occurs.

There is also an issue that we can address now. In both \cite{hsiao} and \cite{lingeng}, the width of the $\rho$ meson is taken constant. However, for a $\rho$ off-shell, the width is energy dependent, as,
\begin{eqnarray}
 \Gamma_\rho(M_\mathrm{inv})=\Gamma_\rho(m_\rho)\frac{m^2_\rho}{M_\mathrm{inv}^2}\left(\frac{q}{q_\mathrm{on}}\right)^3\ ,\label{eq:width}
\end{eqnarray}
with $q=\frac{\lambda^{1/2}(M^2_\mathrm{inv},m^2_\pi,m^2_\pi)}{2M_\mathrm{inv}}\theta(M_\mathrm{inv}-2m_\pi)$\ ,
and $q_\mathrm{on}$ being given by the same expression, but changing $M_\mathrm{inv}$ by $m_\rho$. We can evaluate $M^2_\mathrm{inv}=(P^0-q^0)^2-\vec{q}\,^2$ using the Cauchy integration, since $q^0=\omega_\rho(q)$, for $t_1$ and $q^0=P^0-\omega_\rho(q)$ for $t_2$. We show the results of the consideration of these issues in the next section.
\section{Results}
In what follows, we remove the $t_{\pi^0\eta,\pi^0\eta}(M_\mathrm{inv})$ amplitude since it factorizes outside the integral. What we show below corresponds to taking $Ct_{\pi^0\eta,\pi^0\eta}(M_\mathrm{inv})=1$. 

\begin{figure}
\begin{center}
 \begin{tabular}{c}
 \includegraphics[scale=0.65]{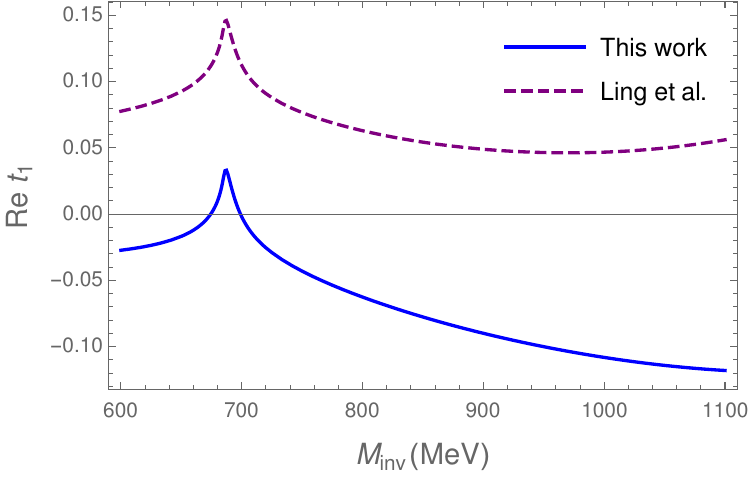}\\ \includegraphics[scale=0.65]{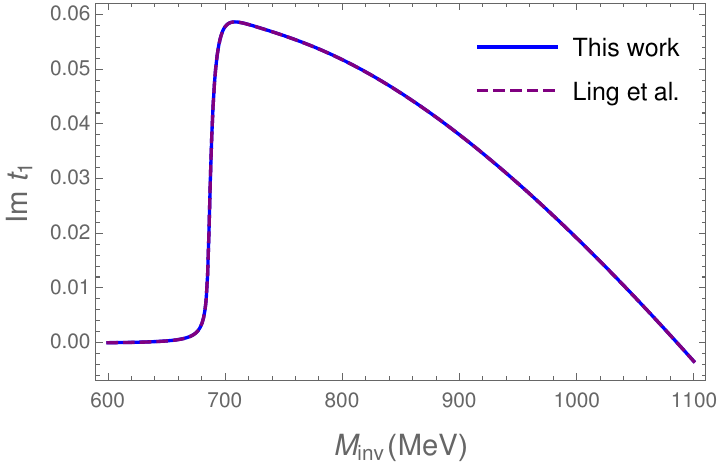}
 \end{tabular}
 \end{center}
 \caption{Real and imaginary parts of the amplitude $t_1$ in Eq.~(\ref{eq:t1}) with and without the $\theta(\cdot)$ function.}
 \label{fig:t1}
\end{figure}
In Fig.~\ref{fig:t1} (upper panel), we show the results of $\mathrm{Re}t_1$ with or without the $\theta(\cdot)$ function using in both cases $\Gamma_\rho$ fixed. We observe a drastic reduction of the amplitude by including the effect of the $\theta(\cdot)$ function tied to the range of the strong interaction. Except in the peak of this amplitude, the consideration of the $\theta(\cdot)$ function even changes the sign of the amplitude. The peak is due to the opening of the $\pi^0\eta$ channel. It is further enhanced by the near presence of a triangle singularity, as one can see applying the rules of \cite{trianus}. However, we should note that this amplitude must be multiplied by $t_{\pi^0\eta,\pi^0\eta}(M_\mathrm{inv})$, which peaks around the $a_0(980)$, and its strength around $700$~MeV is very small. Thus, one should not expect to see much of this structure in the actual experiment.

Next, we show in Fig.~\ref{fig:t1} (lower panel), $\mathrm{Im}t_1$. There are contributions to this imaginary part from two cuts in the diagram, the one placing the $\eta$ and $\rho^+$ on-shell, and the one of the $\eta$ and $\pi^0$ on-shell. We see that below the $\pi^0\eta$ threshold, the imaginary part of $t_1$ is very small and it grows very fast from the threshold on. The largest part of $\mathrm{Im}t_1$ comes from the $\pi^0\eta$ cut. Below threshold, $\mathrm{Im}t_1$ is affected a bit by the $\theta(\cdot)$ function, but this function is $1$ for the $\pi^0\eta$ cut and hence, we see no effect of it in $\mathrm{Im}t_1$.

\begin{figure}
\begin{center}
 \begin{tabular}{c}
 \includegraphics[scale=0.65]{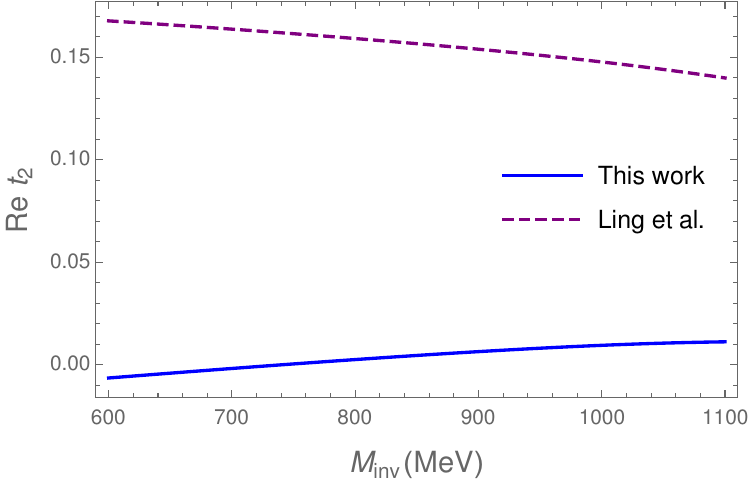}\\ \includegraphics[scale=0.65]{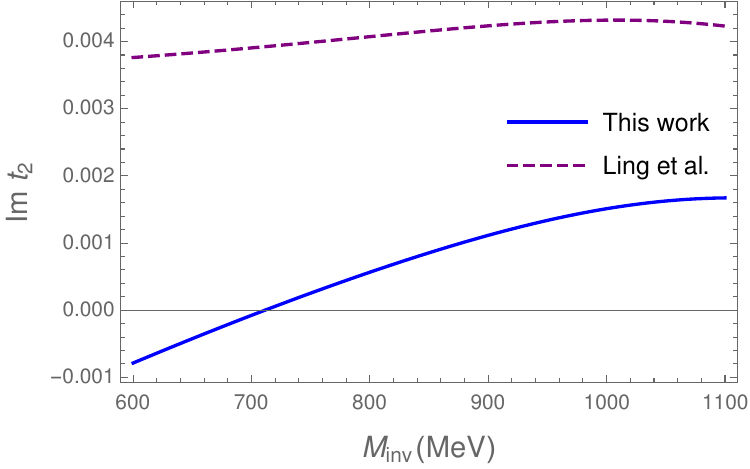}
 \end{tabular}
 \end{center}
 \caption{Real and imaginary parts of the amplitude $t_2$ in Eq.~(\ref{eq:t2}) with and without the $\theta(\cdot)$ function.}
 \label{fig:t2}
\end{figure}
In Fig.~\ref{fig:t2} (upper panel), we show $\mathrm{Re}t_2$, with and without the effect of the $\theta(\cdot)$ function. Once again, we see a strong reduction of this amplitude due to the consideration of the $\theta(\cdot)$ function. We should note that, while $\mathrm{Re}t_1$ and $\mathrm{Re}t_2$ have similar strength without the $\theta(\cdot)$ function, when including it, the strength of $\mathrm{Re}t_2$ becomes smaller than that of $\mathrm{Re}t_1$.

In Fig.~\ref{fig:t2} (lower panel), we show $\mathrm{Im}t_2$ with or without the $\theta(\cdot)$ function. We see that this imaginary part is about one order of magnitude smaller than $\mathrm{Re}t_1$. This is because, in this case, the only source of imaginary part comes from $\eta \rho^+$ being on-shell. The other cut correspondng to placing $\rho^+$ and $\pi^0$ on-shell (see second denominator in Eq.~(\ref{eq:t2})) does not give imaginary part because the process $\pi\to\pi\rho$ is not allowed. In any case, we see that the consideration of the $\theta(\cdot)$ function reduces the strength of the imaginary part (note that we have $\theta(q_\mathrm{max}-|\vec{q}\,^\star|)$, and not, $\theta(q_\mathrm{max}-|\vec{q}\,|)$, although, given its small strength, it does not have any relevant consequence.

\begin{figure}
\begin{center}
 \begin{tabular}{c}
 \includegraphics[scale=0.65]{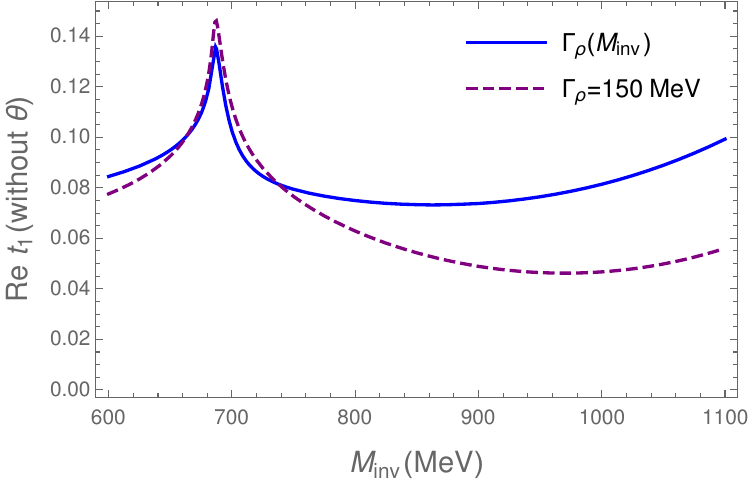}\\ \includegraphics[scale=0.65]{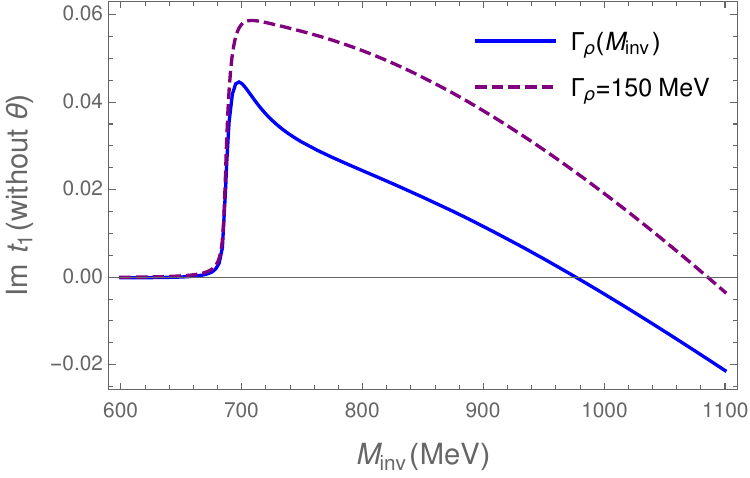}
 \end{tabular}
 \end{center}
 \caption{Real and imaginary parts of the amplitude $t_1$ in Eq.~(\ref{eq:t1}) without the $\theta(\cdot)$ function, taking a $\rho$ width energy dependent or constant, see Eq.~(\ref{eq:width}).}
 \label{fig:width}
\end{figure}
Next, we look into the effect of considering the energy dependence of the $\rho$ width in $\mathrm{Re}t_1$. We see this in Fig.~\ref{fig:width} (upper panel). We show the effects of only this energy dependence, and thus, in both cases, we have removed the effect of the $\theta(\cdot)$ function. In this case, we observe that in the region of $M_\mathrm{inv}\sim 1000$~MeV, where the $a_0(980)$ gives its maximum strength, the consideration of the energy dependence of the $\rho$ width increases the value of $\mathrm{Re}t_1$.

In Fig.~\ref{fig:width} (lower panel), we show the effect of the energy dependent $\rho$ width. In this case, the consideration of the energy dependence reduces the strength of $\mathrm{Im}t_1$. We refrain from showing this effect in $\mathrm{Re}t_2$ and $\mathrm{Im}t_2$, because in this case, $q^0=P^0-\omega_\rho(q)$, and hence, $M^2_{\mathrm{inv}}(\rho)=(P^0-q^0)^2-\vec{q}\,^2=\omega_\rho(q)^2-\vec{q}\,^2=m_\rho^2$. Hence, the use of the on-shell $\Gamma_\rho$ width for this case is justified. 

\begin{figure}
 \centering
 \includegraphics[scale=0.65]{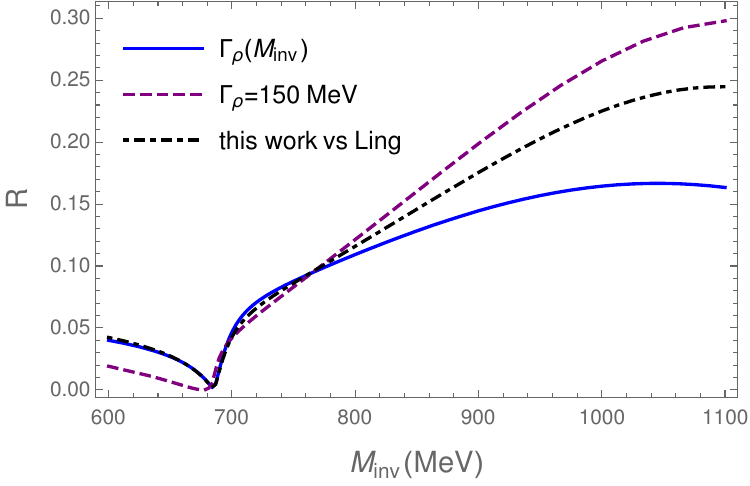}
 \caption{Ratio $R=\vert t_1+t_2\vert^2 $(with $\theta(\cdot)$)$/\vert t_1+t_2\vert^2 $(without $\theta(\cdot)$), using a constant $\rho$ width (continuous line) in both, numerator and denominator, an energy dependent width (dashed line), and, with a constant $\rho$ width in the denominator and an energy dependent width in the numerator (dot-dashed line).}
\end{figure}

Finally, we show the ratio of $R=\vert t_1+t_2\vert^2 $(with $\theta(\cdot)$)$/\vert t_1+t_2\vert^2 $(without $\theta(\cdot)$), considering the elements discussed above. This should give us an idea, by looking around $M_\mathrm{inv}\simeq 1000$~MeV, of the relevance of the corrections in the $D_s\to \pi^+\pi^0\eta$ decay width evaluated in \cite{lingeng}. The continuous curve corresponds to the ratio $R$ with energy dependent width of the $\rho$ in both, numerator and denominator. We observe a drastic reduction of the magnitude $\vert t_1+t_2\vert^2$ from the consideration of the $\theta(\cdot)$ function by a factor of about $0.15$. The dashed curve, where the reduction is smaller than before, corresponds to the same ratio but with a constant width of the $\rho$ meson of $150$~MeV. Finally, the intermediate dot-dashed line, corresponds to putting $\Gamma_\rho$ constant in the denominator with no $\theta(\cdot)$ function, and the energy dependent $\rho$ width in the numerator with the $\theta(\cdot)$ function. This curve shows the combined effect of the two ingredients discused in this work with respect to the calculation of \cite{lingeng}. The reduction factor is about $0.25$ around $1000$~MeV. Note that $\sqrt{0.25}=0.5$, and therefore, one would expect a reduction of about a factor of two in $t_1+t_2$ on the results of \cite{lingeng}, which would still leave this mechanism with a sizeable strength.
 We note that in the dimensional regulization scheme of [7], one can use different subtraction constants in the two-pion one-loop function and in the three-point one-loop function. As a result, is not so surprising to see the reduction by a factor of four. In other words, the method proposed in this work is more consistent in the case where  on-shell UChPT amplitudes are part of the triangle diagram.

\section{Conclusions}    
   We have discussed some important issues concerning triangle loops and how to solve them technically. Concretely, we have addressed the problem of taking into account the range of the interaction which enters in some loop mechanisms that involve one strong interaction transition matrix element. This is usually not done in works that use Feynman diagrammatic techniques to evaluate the loop integral. Another issue that we have addressed is the one of the energy dependent width of unstable particles in the loop, which again is taken constant in the Feynman integral methods. The way we have addressed the problem is by performing the $q^0$ integration of the $d^4q$ integral analytically. This is done with the Cauchy residues method which renders the value of $q^0$ at which the residue has to be evaluated and allows to incorporate the corrections mentioned above in a simple way.  
The method also allows us to have an insight at the convergence of the diagrams when cutoffs or form factors are used, and prevents us from getting spurious contributions related to the poles implicit in some form factors which have no physical meaning.  

     In particular, we have considered the triangle loop that appears in the $D_s^+ \to \rho^+\eta$ decay, followed by $\rho^+ \to \pi^+ \pi^0$ and rescattering of the $\pi^0 \eta$ producing the $a_0(980)$ resonance. The method proposed in this work is particularly suitable when the on-shell unitary chiral amplitudes are used to describe hadron-hadron interactions, such as the $\pi\eta\to\pi\eta$ and $K\bar{K}\to\pi\eta$ interactions. We have investigated independently the effects due to the range of the strong interaction in the $\pi^0\eta$ scattering and the effects of considering the energy dependence of the rho width inside the loop. We find that both effects are important resulting in sizeable changes.  When all effects are considered together, we find a reduction of the strength of the decay width around the $\pi^0\eta$ invariant mass in the $a_0(980)$ region of about a factor $4$. This is sizeable and important to consider if one wishes to have accurate results. Concerning the previous work done in the subject, these effects could reduce the strength of the amplitudes in about a factor of two, which still would make them relevant, but it leaves more room for another mechanism previously claimed, based on internal emission, different than the triangle mechanism proposed as an alternative method to explain the experimental data. While there is room for both mechanisms, we find that the triangle mechanism should have a smaller strength than so far calculated, which calls for a relevant role of the internal emission mechanism in addition. 
     \\
     \section{Acknowledgments}

%a careful reading of the paper and 
R. M. acknowledges support from the CIDEGENT program with Ref. CIDEGENT/2019/015, the Spanish Ministerio de Economia
y Competitividad and European Union (NextGenerationEU/PRTR) by the grant with Ref. CNS2022-136146. 
Li-Sheng Geng is partly supported by the National Natural Science Foundation of China under Grants No.11975041 and No.11961141004. Ming-Zhu Liu acknowledges support from the National Natural Science Foundation of
China under Grant No.12105007. This work is also partly supported by the Spanish Ministerio de Economia y Competitividad (MINECO) and
European FEDER funds under Contracts No. FIS2017-84038-C2-1-P B, PID2020-112777GB-I00, and by Generalitat Valenciana under contract PROMETEO/2020/023. This project has received funding from the European Union Horizon 2020 research
and innovation programme under the program H2020-INFRAIA-2018-1, grant agreement No. 824093 of the STRONG-2020
project.

\bibliography{biblio}

\end{document}